\documentclass[prl,twocolumn,floatfix,showpacs]{revtex4}

\usepackage{amssymb}
\usepackage{amsmath}
\usepackage{amsfonts}
\usepackage{graphicx}
\usepackage{bm}

\begin{document}

\title{Entanglement Observables and Witnesses for Interacting Quantum Spin Systems}

\author{L.-A. Wu}
\affiliation{Chemical Physics Theory Group, Department of Chemistry, and Center for
Quantum Information and Quantum Control, University of Toronto, 80 St.
George St., Toronto, Ontario, M5S 3H6, Canada}
\author{S. Bandyopadhyay}
\altaffiliation[Present address: ]{Institute for Quantum Information Science, 
University of Calgary, Calgary, AB, T2N 1N4, Canada}
\affiliation{Chemical Physics Theory Group, Department of Chemistry, and Center for
Quantum Information and Quantum Control, University of Toronto, 80 St.
George St., Toronto, Ontario, M5S 3H6, Canada}
\author{M. S. Sarandy}
\affiliation{Chemical Physics Theory Group, Department of Chemistry, and Center for
Quantum Information and Quantum Control, University of Toronto, 80 St.
George St., Toronto, Ontario, M5S 3H6, Canada}
\author{D. A. Lidar}
\altaffiliation[Present address: ]{Departments of 
Chemistry and Electrical Engineering, University of
Southern California,  Los Angeles, CA 90089}
\affiliation{Chemical Physics Theory Group, Department of Chemistry, and Center for
Quantum Information and Quantum Control, University of Toronto, 80 St.
George St., Toronto, Ontario, M5S 3H6, Canada}

\begin{abstract}
We discuss the detection of entanglement in interacting quantum spin systems. First, thermodynamic Hamiltonian-based witnesses are computed
for a general class of one-dimensional spin-$1/2$ models. Second, we
introduce optimal bipartite entanglement observables. We show that a bipartite
entanglement measure can generally be associated to a set of independent
two-body spin observables whose expectation values can be used to witness
entanglement. The number of necessary observables is ruled by the symmetries
of the model. Illustrative examples are presented.
\end{abstract}

\maketitle

Entanglement is a striking feature of quantum mechanics, revealing the
existence of non-local correlations among different parts of a quantum
system. Entanglement has been recognized
as an essential resource for quantum information processing \cite{Nielsen:book}. This has provided strong motivation for studies probing for the presence
of \emph{naturally available} entanglement in interacting spin systems \cite{OConnor:01,Arnesen:01Gunlycke:01Wang:01}.
Moreover, the realisation that entanglement can also affect macroscopic
properties (such as the magnetic susceptibility) of bulk solid-state
systems ~\cite{Ghosh:03,Brukner:04a},
has increased the interest in
characterizations of entanglement in terms of \emph{macroscopic
thermodynamical observables}. An observable which can distinguish between
entangled and separable states in a quantum system is called an entanglement
witness~\cite{Terhal:00aTerhal:02Bruss:02}. Several different methods for experimental detection of entanglement using witness operators have been proposed~\cite{Bourennane:04Rahimi:04Stobinska:04}. Entanglement witnesses have recently been obtained in terms of expectation
values of thermodynamical observables such as internal energy and
magnetization~\cite{Brukner:04,Toth:04,Bartlett:04}, and magnetic
susceptibility~\cite{Brukner:04a}.

Our aim in this work is two-fold: first, we find an entanglement witness
for a broad class of interacting spin-$1/2$ particles, thus
generalizing the result of
Refs.~\cite{Brukner:04,Toth:04,Bartlett:04}. This is an entanglement witness for all spin-$1/2$ based
solid-state quantum computing proposals, such as electron spins in quantum dots \cite{Loss:98}
and P donors in Si \cite{Kane:98Vrijen:00}. While this approach is very general, its
drawback is that it is sub-optimal, in the sense that it does not detect all
entangled states. In contrast, in the second part of this work, we introduce
the concept of an optimal bipartite entanglement observable. This allows
us to construct \emph{optimal} bipartite-entanglement witnesses for qubit
systems. The essential idea here is to directly relate bipartite
entanglement measures and the expectation value of spin observables~\cite{Wang:02a}. 

\textit{Hamiltonian-based entanglement witnesses}.--- An important class of
spin-based solid-state quantum computing proposals is approximately governed
by \emph{diagonal} exchange interactions (involving only 
$\sigma_i^{\alpha}\sigma_j^{\alpha}$ terms, where $\alpha \in \{x,y,z \}$ 
and $\sigma_i^{\alpha}$ is the Pauli matrix for spin $i$)~\cite{Loss:98,Kane:98Vrijen:00}. 
However, spin-orbit coupling introduces off-diagonal
terms into the exchange Hamiltonian \cite{Kavokin:01}. In this case
previous results concerning Hamiltonian-based entanglement witnesses~\cite{Toth:04,Brukner:04,Bartlett:04} 
do not apply, since they are restricted to the diagonal case. Here we construct an appropriately 
generalized entanglement witness.

The most general Hamiltonian describing $N$ nearest-neighbor coupled spin-$1/2$ particles in
1D is of the form $H=\sum_{i}\sum_{\alpha ,\beta \in
\{x,y,z\}}g_{i,i+1}^{\alpha \beta }\sigma _{i}^{\alpha }\sigma _{i+1}^{\beta }$,
where 
$g_{i,i+1}^{\alpha \beta }=(g_{i,i+1}^{\beta \alpha })^{\ast }$. There are thus
nine independent parameters for each pair of spins $i,i+1$. It is convenient
to re-express $H$ in terms of a scalar part and symmetric and anti-symmetric
parts. In addition we allow for the presence of a global external magnetic field $\mathbf{B}$:
$H =-\mathbf{B}\cdot \sum_{i=1}^{N}{\bm \sigma}_{i}+\sum_{i=1}^{N}\sum_{
\alpha =x,y,z}J_{\alpha }\sigma _{i}^{\alpha }\sigma _{i+1}^{\alpha } \\
+\sum_{i=1}^{N}\mathbf{A}\cdot ({\bm \sigma}_{i}\times {\bm \sigma}_{i+1}
)+\sum_{i=1}^{N}(\mathbf{C}\cdot {\bm \sigma}_{i})(\mathbf{C}\cdot {\bm \sigma}_{i+1})
$, where ${\bm \sigma}_{i}=(\sigma _{i}^{x},\sigma _{i}^{y},\sigma _{i}^{z})$, 
$J_{\alpha }$ are exchange coupling constants, and we assume periodic
boundary conditions (${\bm \sigma}_{N+1}\equiv {\bm \sigma}_{1}$). The
anisotropic term involving $\mathbf{A}$ (the Dzyaloshinskii-Moriya vector in
solid-state physics) typically arises due to spin-orbit
coupling; $|\mathbf{A}|/J$ has been estimated to be in the range $0.01-0.8$ in coupled quantum dots in 
GaAs \cite{Kavokin:01}. The vector $\mathbf{C}$ can arise also due
to dipole-dipole coupling and other sources.

We now derive a thermodynamical entanglement witness for a system governed
by $H$. Let $J=\max_{\alpha
}\{|J_{\alpha }|\}$, $A=\max_{\alpha }\{|A_{\alpha }|\}$. Let $\langle
X\rangle \equiv \mathrm{Tr}(\rho X)$, with $\rho $ the system density
matrix. Let $u=\left\langle H\right\rangle /N$ be the per-spin internal
energy and $\mathbf{m}=(m_{x},m_{y},m_{z})$ the magnetization vector with
components $m_{\alpha }=\sum_{i=1}^{N}\left\langle \sigma _{i}^{\alpha
}\right\rangle /N$. Then $H$ yields:

\begin{eqnarray*}
&N& \!\!\! \left\vert u+\mathbf{B}\cdot \mathbf{m}\right\vert \leq \left\vert
\sum_{i=1}^{N}\sum_{\alpha =x,y,z}J_{\alpha }\left\langle \sigma
_{i}^{\alpha }\sigma _{i+1}^{\alpha }\right\rangle \right\vert \\
&&+\left\vert \sum_{i=1}^{N}\left\langle \mathbf{A}\cdot ({\bm \sigma}_{i}
\times {\bm \sigma}_{i+1})\right\rangle \right\vert +\left\vert
\sum_{i=1}^{N}\left\langle (\mathbf{C\cdot }{\bm \sigma}_{i})(\mathbf{
C\cdot }{\bm \sigma}_{i+1})\right\rangle \right\vert .
\end{eqnarray*}
Consider an arbitrary separable density matrix $\rho =\sum_{k}p_{k}\rho
_{k}^{1}\otimes \rho _{k}^{2}\otimes ...\otimes \rho _{k}^{N}$, where $
\sum_{k}p_{k}=1$ and all $p_{k}\geq 0$. It has been shown that for such $
\rho $, using the easily verified facts $\langle \sigma _{i}^{\alpha }\sigma
_{i+1}^{\beta }\rangle =\left\langle \sigma _{i}^{\alpha }\right\rangle
\langle \sigma _{i+1}^{\beta }\rangle $ and $\sum_{\alpha
=x,y,z}\left\langle \sigma _{i}^{\alpha }\right\rangle ^{2}\leq 1$, and the
Cauchy-Schwarz (CS) inequality $|\sum_{i}a_{i}b_{i}|\leq
(\sum_{i}a_{i}^{2}\sum_{j}b_{j}^{2})^{1/2}$, that $\left\vert
\sum_{i=1}^{N}J_{x}\left\langle \sigma _{i}^{x}\sigma
_{i+1}^{x}\right\rangle +J_{y}\left\langle \sigma _{i}^{y}\sigma
_{i+1}^{y}\right\rangle +J_{z}\left\langle \sigma _{i}^{z}\sigma
_{i+1}^{z}\right\rangle \right\vert \leq NJ$ \cite{Toth:04,Brukner:04}. We
therefore obtain bounds for the remaining two terms. Let $x_{i}\equiv
\left\langle \sigma _{i}^{x}\right\rangle $, $(xy)_{i}\equiv \left\langle
\sigma _{i}^{x}\sigma _{i+1}^{y}\right\rangle $, etc. Using again the above
facts and the CS inequality we have

\begin{eqnarray*}
&|& \!\!\! \sum_{i=1}^{N}\left\langle \mathbf{A}\cdot
  ({\bm \sigma}_{i}\times {\bm \sigma}_{i+1})\right\rangle | \leq \sum_{i=1}^{N}|\left\langle \mathbf{A}
\cdot ({\bm \sigma}_{i}\times {\bm \sigma}_{i+1})\right\rangle | \\
&& \!\!\! \leq 2A \sum_{i=1}^{N} \left\vert
(yz)_{i}+(zx)_{i}+(xy)_{i}\right\vert \\
&& \!\!\! = 2A\sum_{i=1}^{N}\left\vert
y_{i}z_{i+1}+z_{i}x_{i+1}+x_{i}y_{i+1}\right\vert \\
&& \!\!\! \leq
2A\sum_{i=1}^{N}(\sum_{\alpha }\left\langle \sigma _{i}^{\alpha
}\right\rangle ^{2})^{1/2}(\sum_{\alpha }\left\langle \sigma _{i+1}^{\alpha
}\right\rangle ^{2})^{1/2} \leq 2NA.
\end{eqnarray*}
Note that if we assume no symmetry breaking, i.e., $\left\langle \sigma
_{i}^{\alpha }\right\rangle \equiv \left\langle \sigma ^{\alpha
}\right\rangle $, then in fact $|\sum_{i=1}^{N}\left\langle \mathbf{A}\cdot (
{\bm \sigma}_{i}\times {\bm \sigma}_{i+1})\right\rangle |=0$.

We now obtain an upper bound for the third term. In the standard Bloch-sphere
parametrization for the individual spin density matrices we have: $\rho =\frac{1}{2}
(I+\mathbf{n}\cdot {\bm \sigma})$, where $|\mathbf{n}|\leq 1$. Then $
\left\langle \mathbf{C}\cdot {\bm \sigma}\right\rangle =\mathbf{C}\cdot 
\mathbf{n}\leq \left\vert \mathbf{C}\right\vert |\mathbf{n}|\leq \left\vert 
\mathbf{C}\right\vert $. Therefore,

\begin{eqnarray*}
&|& \!\!\! \sum_{i=1}^{N}\left\langle (\mathbf{C}\cdot {\bm \sigma}_{i})(
\mathbf{C}\cdot {\bm \sigma}_{i+1})\right\rangle | \leq
\sum_{i=1}^{N}\left\vert \left\langle (\mathbf{C}\cdot {\bm \sigma}_{i})(
\mathbf{C}\cdot {\bm \sigma}_{i+1})\right\rangle \right\vert \\
&&=\sum_{i=1}^{N}\left\vert \left\langle \mathbf{C}\cdot {\bm \sigma}_{i}
\right\rangle \right\vert \left\vert \left\langle \mathbf{C}\cdot
{\bm \sigma}_{i+1}\right\rangle \right\vert \leq N\left\vert
\mathbf{C}\right\vert ^{2}.
\end{eqnarray*}
These upper bounds combine to yield the entanglement witness:

\begin{equation}
W\equiv \left\vert u+\mathbf{B}\cdot \mathbf{m}\right\vert /(J+2A+\left\vert 
\mathbf{C}\right\vert ^{2}).
\end{equation}
The numerator consists of macroscopic, observable quantities. The
denominator consists of material parameters. We have seen that separability
implies $W\leq 1$. Therefore if $W>1$ the system is entangled. When $J_\alpha \equiv J$ and the anistropic terms in $H$ are
entirely due to the spin-orbit interaction, it is possible to relate
$H$ to the isotropic Heisenberg Hamiltonian via a unitary
transformation \cite{Kavokin:01,WuLidar:03}. Applying this transformation to the 
examples of entangled states that are detected by $W$ in the case $\mathbf{A}
=\mathbf{C}=\mathbf{0}$, found in
Refs.~\cite{Toth:04,Brukner:04}, yields examples of non-trivial
entangled states detected by $W$ when $\mathbf{A},\mathbf{C} \neq \mathbf{0}$.
The importance of the witness $W$ is that it is directly applicable to a
wide class of spin-$1/2$ based solid-state quantum computing proposals
\cite{Loss:98,Kane:98Vrijen:00}, where the effect of spin-orbit
coupling is known to be 
non-negligible \cite{Kavokin:01}.

\textit{Spin-based entanglement witnesses}.--- Let us turn now to the
construction of optimal bipartite-entanglement witnesses, based on spin
observables. Consider a general two-body observable $\widehat{R}=\sum
R_{\alpha \beta \gamma \delta }(ij)\left\vert \alpha \right\rangle
_{i}\left\vert \beta \right\rangle _{j}\left\langle \gamma \right\vert
_{i}\left\langle \delta \right\vert _{j}$, where $\{\left\vert \alpha
\right\rangle _{i}\}$ is a basis for the Hilbert space, $i,j$ enumerate $d$
-level systems, and $\alpha ,\beta ,\gamma ,\delta \in \{0,1,...,d-1\}$. The
expectation value of $\widehat{R}$ can generally be written as~\cite{WuSarandyLidar:04} 
\begin{equation}
\langle \,\widehat{R}\,\rangle =\sum_{ij}{\text{Tr}}(\mathbf{R}(ij)\mathbf{
\rho }^{ij}),  \label{ev}
\end{equation}
where $\mathbf{R}(ij)$ are $d\times d$ matrices with elements $R_{\alpha
\beta \gamma \delta }(ij)$ and $\rho ^{ij}$ is the two-body reduced density
matrix. Eq.~(\ref{ev}) holds for any mixed state and for any $d$. Here we
are especially interested in $d=2$, i.e., the qubit case. We then use the
standard basis $\left\{ \left\vert 00\right\rangle ,\left\vert
01\right\rangle ,\left\vert 10\right\rangle ,\left\vert 11\right\rangle
\right\} $ for any pair $(i,j)$ of spins, and denote $\rho
_{11}^{ij}=\left\langle 0_{i}0_{j}\right\vert {\hat{\rho}}^{ij}\left\vert
0_{i}0_{j}\right\rangle $, $\rho _{12}^{ij}=\left\langle
0_{i}0_{j}\right\vert {\hat{\rho}}^{ij}\left\vert 0_{i}1_{j}\right\rangle $,
etc. For an operator $\widehat{R}$ displaying a constant interaction between
nearest-neighbor particles, the non-vanishing matrices $\mathbf{R}(ij)$ are
given by $\mathbf{R}(i,i+1)\equiv \mathbf{R}$ $(\forall \,i)$. Moreover, if
translation invariance is assumed in the system then $\rho ^{i,i+1}\equiv
\rho $. Hence we can rewrite Eq.~(\ref{ev}) as 
\begin{equation}
\langle \,\widehat{\mathcal{R}}\,\rangle ={\text{Tr}}(\mathbf{R}\mathbf{\rho 
}),  \label{evti}
\end{equation}
where $\widehat{\mathcal{R}}=\widehat{R}/N$, with $N$ the number of
nearest-neighbor pairs in the system.

A convenient bipartite entanglement measure is the
negativity~\cite{Vidal:02a}, ranging from $0$ (no entanglement) to $1$
(maximal entanglement), defined as follows: 
\begin{equation}
\mathcal{N}(\rho )=2\,\max (0,-\min_{\alpha }(\mu _{\alpha })),
\label{negdef}
\end{equation}
where $\mu _{\alpha }$ are the eigenvalues of the partial transpose $\rho
^{T_{A}}$ of the two-particle reduced density operator $\rho $, given by $
\left\langle \alpha \beta \right\vert \rho ^{T_{A}}\left\vert \gamma \delta
\right\rangle =\left\langle \gamma \beta \right\vert \rho \left\vert \alpha
\delta \right\rangle $. We denote the lowest eigenvalue of $\rho ^{T_{A}}$
by $\mu _{m}$ which, for composite systems of dimensions $2\times 2$ and $
2\times 3$, is non-negative if and only if the state is separable~\cite
{Peres:96,Horodecki:96}. Thus, from Eq.~(\ref{negdef}), one can see that
separability implies vanishing negativity~\cite{Wang:02a}.

The eigenvalue $\mu _{m}$, which is the key object of our framework, is
generally a non-linear function of the matrix elements of the density
operator. However, let us first consider the linear case, which occurs for
several interesting quantum spin systems, as will be illustrated later. In
this case, we can directly relate $\mu _{m}$ to a single observable which
plays the role of an entanglement witness. Indeed, assume that $\mu
_{m}=\sum_{a,b=1}^{4}f_{ab}\rho _{ab}$, where $f_{ab}$ are constants. Then,
defining the matrix elements of $\widehat{\mathcal{R}}$ as $R_{ab}=f_{ba}$
we obtain 
\begin{equation}
\mu _{m}={\text{Tr}}(\mathbf{R}\rho )=\langle \,\widehat{\mathcal{R}}
\,\rangle .  \label{flinearmu}
\end{equation}
Therefore, the observable $\widehat{\mathcal{R}}$ directly detects the
existence of bipartite entanglement in the system. $\langle \,\widehat{
\mathcal{R}}\,\rangle <0$ implies bi-partite entanglement, while otherwise
the state is separable.

Eq.~(\ref{flinearmu}) can also be established, formally, for those cases
where $\mu _{m}$ depends non-linearly on $\rho $. Indeed, from Eq.~(\ref
{evti}) it follows that 
\begin{eqnarray}
\langle \,\widehat{\mathcal{R}}\,\rangle ={\text{Tr}}(\mathbf{R}^{T_{A}}\mathbf{\rho }^{T_{A}}) =\sum_{\alpha =1}^{4}\left\{ W\mathbf{R}^{T_{A}}W^{\dagger }\right\}
_{\alpha \alpha }\mu _{\alpha },  \label{negob}
\end{eqnarray}
where the matrix $\mathbf{R}^{T_{A}}$ is defined through $\left\langle
\alpha \beta \right\vert \mathbf{R}^{T_{A}}\left\vert \gamma \delta
\right\rangle =\left\langle \gamma \beta \right\vert \mathbf{R}\left\vert
\alpha \delta \right\rangle $, $W$ is a unitary matrix which diagonalizes $
\mathbf{\rho }^{T_{A}}$ (note that $\mathbf{\rho }^{T_{A}}$ is Hermitian 
\cite{Peres:96}), and the $\{\mu _{\alpha }\}$ denote the four eigenvalues
of $\mathbf{\rho }^{T_{A}}$. If we choose $\widehat{\mathcal{R}}$ such that 
\begin{equation}
W\mathbf{R}^{T_{A}}W^{\dagger }={\text{diag}}(e_{1},e_{2},e_{3},e_{4}),
\label{negcond}
\end{equation}
where $e_{\alpha }=1$ for the value of $\alpha $ such that $\mu _{\alpha
}=\mu _{m}$ and $e_{\alpha }=0$ for the other ones, then as desired $\langle
\,\widehat{\mathcal{R}}\,\rangle =\mu _{m}$. Hence the expectation value of $
\widehat{\mathcal{R}}$ can be used in general as a criterion of
separability. However, in the non-linear case, if we define $\widehat{
\mathcal{R}}$ through Eq.~(\ref{negcond}), $\widehat{\mathcal{R}}$ itself
will be a function of the density matrix (since $W$ is). One would then need
to measure a complete set of observables to find $\,\widehat{\mathcal{R}}$,
which just corresponds to quantum state tomography
\cite{Nielsen:book}. We show below that in the presence of symmetries the number of measurements
required to construct $\widehat{\mathcal{R}}$ can be drastically reduced.

\textit{XYZ spin chain in a magnetic field}.--- In order to provide an
example of spin-based witnesses in quantum spin chains, let us consider a
parametric family of spin Hamiltonians $H=\sum_{i=1}^{N}H_{i,i+1}(
{\bm \theta}_{i},{\bm \theta}_{i+1})$, where $
H_{i,i+1}({\bm \theta}_{i},{\bm \theta}_{i+1})=U(
{\bm \theta}_{i})U({\bm \theta}_{i+1})H_{i,i+1}U^{
\dagger }({\bm \theta}_{i+1})U^{\dagger }({\bm \theta}_{i})$, $U({\bm \theta}_{i})=\exp (i{\bm \theta}
_{i}\cdot {\bm \sigma }_{i})$, and 
\begin{eqnarray}
H_{i,i+1} &=&J_{x}\sigma _{i}^{x}\sigma _{i+1}^{x}+J_{y}\sigma
_{i}^{y}\sigma _{i+1}^{y}+J_{z}\sigma _{i}^{z}\sigma _{i+1}^{z}  \notag \\
&&+J_{xy}\sigma _{i}^{x}\sigma _{i+1}^{y}+J_{yx}\sigma _{i}^{y}\sigma
_{i+1}^{x}+h\sigma _{i}^{z},  \label{Hgeneral}
\end{eqnarray}
with periodic boundary conditions assumed, i.e. $\sigma _{N+1}^{\alpha
}=\sigma _{1}^{\alpha }$. Observe that a large class of one-dimensional spin
models is covered by the Hamiltonian~(\ref{Hgeneral}). This family of
Hamiltonians obeys the constraint $\left[ H,\sigma _{i}^{z}\sigma _{i+1}^{z}
\right] =0$ and belongs to the subalgebra $su(2)\oplus su(2) \subset su(4)$. Defining $T_{i}^{\pm }\equiv \frac{1}{2}\left( \sigma
_{i}^{x}\sigma _{i+1}^{x}\pm \sigma _{i}^{y}\sigma _{i+1}^{y}\right) $, $
R_{i}^{\pm }\equiv \frac{1}{2}\left( \sigma _{i}^{x}\sigma _{i+1}^{y}\pm
\sigma _{i}^{y}\sigma _{i+1}^{x}\right) $, $Z_{i}^{\pm }\equiv \frac{1}{2}
(\sigma _{i}^{z} \pm \sigma _{i+1}^{z})$, one of the $su(2)$ terms is generated
by the set of operators $\{T_{i}^{+},R_{i}^{-},Z_{i}^{-}\}$ (respectively
with coefficients $J_{x}+J_{y}$, $J_{xy}-J_{yx}$, and $h$, in $H$) and
preserves the two dimensional subspace spanned by $\left\{ \left\vert
01\right\rangle ,\left\vert 10\right\rangle \right\} $. The other $su(2)$ is
generated by $\{T_{i}^{-},R_{i}^{+},Z_{i}^{+}\}$ (respectively with
coefficients $J_{x}-J_{y}$, $J_{xy}+J_{yx}$, and $h$, in $H$) and preserves
the other two dimensional subspace spanned by $\left\{ \left\vert
00\right\rangle ,\left\vert 11\right\rangle \right\} $. The Hamiltonian $
H_{i,i+1}({\bm \theta}_{i},{\bm \theta}_{i+1})$ has
the same entanglement properties as $H_{i,i+1}$ since they are connected
through local unitary transformations. Therefore, assuming that the initial
state has the same symmetry as the Hamiltonian (no symmetry breaking), the
nearest-neighbor reduced density matrix for an arbitrary mixed state reads,
in the standard $\{\left\vert 00\right\rangle ,\left\vert 01\right\rangle
,\left\vert 10\right\rangle ,\left\vert 11\right\rangle \}$ basis 
\begin{equation}
\rho =\left[ 
\begin{array}{cccc}
a & 0 & 0 & y \\ 
0 & b & z & 0 \\ 
0 & z^{\ast } & c & 0 \\ 
y^{\ast } & 0 & 0 & d
\end{array}
\right] .  \label{dmgeneral}
\end{equation}
Positivity of $\rho $ implies that $ad\geq |y|^{2}$ and $bc\geq |z|^{2}$,
with $a,b,c,d\geq 0$. Computing the eigenvalues of $\rho ^{T_{A}}$ leads to
two independent possibilities for the lowest eigenvalue $\mu _{m}$: 
\begin{eqnarray}
\mu _{m}^{(1)} &=&(a+d-\sqrt{(a-d)^{2}+4\left\vert z\right\vert ^{2}})/2,
\label{case1} \\
\mu _{m}^{(2)} &=&(b+c-\sqrt{(b-c)^{2}+4\left\vert y\right\vert ^{2}})/2.
\label{case2}
\end{eqnarray}
The condition for entanglement $\mu _{m}<0$, together with positivity of $
\rho $, yields the restrictions 
\begin{eqnarray}
\mu _{m}^{(1)} &<&0\Rightarrow ad<bc,  \label{condcase1} \\
\mu _{m}^{(2)} &<&0\Rightarrow bc<ad.  \label{condcase2}
\end{eqnarray}
Note that, since $\mu _{m}$ is nonlinear in the density matrix elements, the
resulting witness $\widehat{\mathcal{R}}$, obtained from Eqs.~(\ref{negob})
and~(\ref{negcond}), will be density matrix-dependent. Indeed, for $\mu
_{m}=\mu _{m}^{(1)}$, 
\begin{eqnarray}
  \widehat{\mathcal{R}}&=&{\rm Im}(f)R^{-}+{\rm Re}(f)T^{+}
  -\frac{1}{2}\frac{(a-d)}{\sqrt{(a-d)^{2}+4|z|^{2}}}Z^{+} \notag \\
&+& \!\!\! \frac{1}{4}\left(
I\otimes I+\sigma ^{z}\otimes \sigma ^{z}\right) ,  \label{r1}
\end{eqnarray}
where $f\equiv -z/\sqrt{(a-d)^{2}+4|z|^{2}}$, and $R^{-},T^{+},Z^{+}$ were
defined above. It is seen from this result that the state-dependence can be
removed if the following constraints are obeyed: $a=d$ and $z$ is either
real or imaginary. These constraints are obeyed, e.g., for the isotropic
Heisenberg model (see below). In this case we need to measure just one
observable in order to determine the entanglement properties of the system.
Analyzing the second possibility, i.e. $\mu _{m}=\mu _{m}^{(2)}$, we obtain
\begin{eqnarray}
  \widehat{\mathcal{R}}&=&-{\rm Im}(g)R^{+}+ {\rm Re}(g)T^{-}
  -\frac{1}{2}\frac{(b-c)}{\sqrt{(b-c)^{2}+4|y|^{2}}}Z^{-} \notag \\
&+&\frac{1}{4}\left(
I\otimes I-\sigma ^{z}\otimes \sigma ^{z}\right) ,  \label{r2}
\end{eqnarray}
where $g\equiv -y/\sqrt{(b-c)^{2}+4|y|^{2}}$, and $R^{+},T^{-},Z^{-}$ were
defined above. Similarly, $\widehat{\mathcal{R}}$ is state-independent in
Eq.~(\ref{r2}) for $b=c$ and $y$ either real or imaginary. An example of
this case is given by the transverse field Ising model (see below). But even
in the case of the rather general Hamiltonian (\ref{Hgeneral}), it is clear
that instead of full-scale quantum state tomography, it suffices to measure
the elements $\{a,d,z\}$ or $\{b,c,y\}$, in order to construct the witness
operator $\widehat{\mathcal{R}}$.

\textit{Heisenberg model}.--- Let us consider the constraints $
J_{x}=J_{y}=J_{z}\equiv J>0$ and $J_{xy}=J_{yx}=h=0$ in
Eq.~(\ref{Hgeneral}). Then we have the antiferromagnetic Heisenberg
chain, whose Hamiltonian reads 
$H=J\sum_{i=1}^{N}\left( \sigma _{i}^{x}\sigma _{i+1}^{x}+\sigma
_{i}^{y}\sigma _{i+1}^{y}+\sigma _{i}^{z}\sigma _{i+1}^{z}\right)$.
$H$ is invariant under cyclic spin translations and has $SU(2)$
symmetry, with $H$ commuting with the total spin components $\sum_{i}\sigma
_{i}^{\alpha }$, $\alpha \in \{x,y,z\}$. The elements of $\rho $ in
Eq.~(\ref{dmgeneral}) then obey further constraints, namely, $y=0$,
$z=z^{\ast }<0$, $a=d=1+\langle \sigma _{i}^{z}\sigma _{i+1}^{z}\rangle $, and $b=c=1-\langle
\sigma _{i}^{z}\sigma _{i+1}^{z}\rangle $, with $\langle \sigma
_{i}^{z}\sigma _{i+1}^{z}\rangle \leq 0$~\cite{OConnor:01,Wang:02a}(a). It then follows from Eq.~(\ref{case2})
that the eigenvalue $\mu _{m}^{(2)}$ is always non-negative, whence
entanglement is determined by the eigenvalue $\mu _{m}^{(1)}$, given
by Eq.~(\ref{case1}). Thus, the witness comes from the observable in
Eq.~(\ref{r1}), 
which becomes $\widehat{\mathcal{R}}=(\sigma _{i}^{x}\sigma
_{i+1}^{x}+\sigma _{i}^{y}\sigma _{i+1}^{y}+\sigma _{i}^{z}\sigma
_{i+1}^{z}+I)/4$ (for any site $i$). This yields the entanglement witness $
\langle \,\widehat{\mathcal{R}}\,\rangle =\frac{1}{4}\left( u/J+1\right) $,
where we have used that, due to the translation invariance and $SU(2)$
symmetry, the correlation functions satisfy the relations: $\langle \sigma
_{i}^{x}\sigma _{i+1}^{x}\rangle =\langle \sigma _{i}^{y}\sigma
_{i+1}^{y}\rangle =\langle \sigma _{i}^{z}\sigma _{i+1}^{z}\rangle =u/(3J)$~
\cite{Wang:02a}(a). Remarkably, this spin-based witness is (upto an
irrelevant prefactor) precisely the Hamiltonian-based entanglement witness
found in Ref.~\cite{Brukner:04} (and in our generalized Hamiltonian-based
result above). Since our spin-based approach is optimal for bi-partite
entanglement ($\langle \,\widehat{\mathcal{R}}\,\rangle $ is essentially the
negativity), this is a proof that the Hamiltonian-based witness \cite{Brukner:04} detects \emph{all} bi-partite entangled states.

\begin{figure}[th]
\centering {\includegraphics[angle=0,scale=0.32]{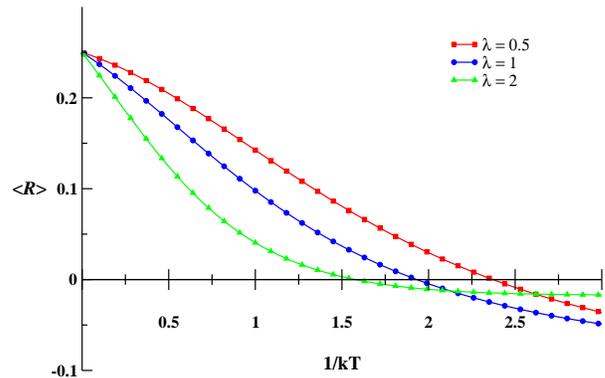}}
\caption{Witness for the transverse field Ising model. The
  entanglement-separability transition temperature is indicated by the
  intersection with the horizontal axis.}
\label{f1}
\end{figure}

\textit{Transverse Field Ising model}.--- As a final example we analyze the
ferromagnetic one-dimensional Ising chain in the presence of a transverse
magnetic field. This model corresponds to taking $J_{x}=-\lambda J$ and $h=-J
$ in Eq.~(\ref{Hgeneral}), with $J>0$ and all the other couplings vanishing: $
H=-J\sum_{i=1}^{N}\left( \lambda \sigma _{i}^{x}\sigma _{i+1}^{x}+\sigma
_{i}^{z}\right) $. Then $
y=y^{\ast }$ ($Z_{2}$ symmetry of $H$) and $b=c$ (translation symmetry
of $H$)~\cite{WuSarandyLidar:04}. From the analysis of the thermal correlation
functions, which can be obtained~analytically \cite{Barouch:71}, it can be
shown that, in contrast to the Heisenberg case, entanglement is now
determined by the eigenvalue $\mu _{m}^{(2)}$ in Eq.~(\ref{case2}). From
Eq.~(\ref{r2}), our spin-based entanglement observable is then $\widehat{
\mathcal{R}}=-\frac{1}{4}(\sigma _{i}^{x}\sigma _{i+1}^{x}-\sigma
_{i}^{y}\sigma _{i+1}^{y}+\sigma _{i}^{z}\sigma _{i+1}^{z}-1)$ for any site $
i$.
The expectation value of this observable can be determined from
measurements of the nearest-neighbor spin-spin correlations $\langle \sigma
_{i}^{\alpha }\sigma _{i+1}^{\alpha }\rangle $. This can be done,
e.g., by inelastic neutron scattering
\cite{Hohenberg:74}. We note that it was shown in
  Ref.~\cite{Brukner:04a} that spin-spin correlation functions can act as 
entanglement witnesses in bulk solids. Our witness operator
$\widehat{\mathcal{R}}$ is optimal, so that it can, moreover, yield precise macroscopic predictions. For instance, we
plot in Fig.~\ref{f1} the witness as a function of $\beta =1/kT$ for several
values of $\lambda $, where $k$ is the Boltzmann constant and $T$ is the
temperature. From this figure, we can obtain the exact critical temperature $T_c$
for the entanglement-separability transition. For $\lambda =1$, we have $
\beta_c \approx 1.93$ or $kT_c\approx 0.51$ (in units such that $J=1$). This
temperature can be compared to the value $0.41$ obtained in Ref.~\cite{Toth:04} by using an energy witness. The reason for the small difference
is that the energy witness of Ref.~\cite{Toth:04} is derived from an
entanglement bound and, despite being a good approximation, neglects some
entangled states which are detected by
$\widehat{\mathcal{R}}$. 

%\textit{Conclusions.---} We discussed entanglement witnesses for a large
%class of spin models. The witnesses were constructed by computing
%expectation values of the Hamiltonian operator and also by a second
%approach, given by spin-based observables. To this end, we introduced a set
%of spin operators whose expectation values characterize two-particle
%entanglement. These expectation values can be significantly simplified in
%the presence of symmetries. In particular, for cases where the bipartite
%entanglement measure is linear in the density matrix elements, only a single
%spin observable was shown to be necessary to characterize entanglement.
%Generalizations of our results for general many-body systems are left for
%future research.

\textit{Acknowledgments.---} We gratefully acknowledge financial support
from CNPq-Brazil (to M.S.S.), and the Sloan Foundation, PREA and NSERC (to
D.A.L.).

%\bibliographystyle{/nfs/theory/u0/dlidar/revtex/prsty}
%\bibliography{/nfs/theory/u0/dlidar/articles/bib}

\end{document}